\documentclass{aa} 

\usepackage{txfonts}
\usepackage{graphicx}
\usepackage[english]{babel}
\usepackage{natbib}
\usepackage[pdftex]{hyperref}
\bibpunct{(}{)}{;}{a}{}{,}

\begin{document}

\title{Long-term luminosity behavior of 14 ultracompact X-ray binaries}

\author{
L.~M.~van~Haaften \inst{\ref{radboud}} \and
R.~Voss \inst{\ref{radboud}} \and
G.~Nelemans \inst{\ref{radboud},\ref{leuven}}
}

\institute{
Department of Astrophysics/ IMAPP, Radboud University Nijmegen, P.O. Box 9010, 6500 GL Nijmegen, The Netherlands, \email{L.vanHaaften@astro.ru.nl} \label{radboud} \and
Institute for Astronomy, KU Leuven, Celestijnenlaan 200D, 3001 Leuven, Belgium \label{leuven}
}

\abstract{X-ray binaries are usually divided in persistent and transient sources. For ultracompact X-ray binaries (UCXBs), the mass transfer rate is expected to be a strong function of orbital period, predicting persistent sources at short periods and transients at long periods.} 
{For 14 UCXBs including two candidates, we investigate the long-term variability and average bolometric luminosity with the purpose of learning how often a source can be expected to be visible above a given luminosity, and we compare the derived luminosities with the theoretical predictions.} 
{We use data from the \emph{RXTE} All-Sky Monitor because of its long-term, unbiased observations. Many UCXBs are faint, i.e., they have a count rate at the noise level for most of the time. Still, information can be extracted from the data, either by using only reliable data points or by combining the bright-end variability behavior with the time-averaged luminosity.} 
{Luminosity probability distributions show the fraction of time that a source emits above a given luminosity. All UCXBs show significant variability and relatively similar behavior, though the time-averaged luminosity implies higher variability in systems with an orbital period longer than $40$ min.} 
{There is no large difference in the statistical luminosity behavior of what we usually call persistent and transient sources. UCXBs with an orbital period below $\sim \! 30$ min have a time-averaged bolometric luminosity that is in reasonable agreement with estimates based on the theoretical mass transfer rate. Around $40$ min the lower bound on the time-averaged luminosity is similar to the luminosity based on the theoretical mass transfer rate, suggesting these sources are indeed faint when not detected. Above $50$ min some systems are much brighter than the theoretical mass transfer rate predicts, unless these systems have helium burning donors or lose additional angular momentum.} 

\keywords{X-rays: binaries -- methods: data analysis -- stars: variables: general}
\authorrunning{van~Haaften, Voss \& Nelemans}

\maketitle

\section{Introduction}

Ultracompact X-ray binaries (UCXBs) are low-mass X-ray binaries characterized by an orbital period of less than one hour. Given that the donor stars in these binary systems fill their Roche lobe, they must be white dwarfs or helium burning stars, since only these types of stars have an average density that corresponds to such a short orbital period \citep{paczynski1981,nelson1986}. The accretors can be neutron stars or black holes, though the latter have not been identified yet.

Historically, X-ray sources including UCXBs have been categorized as persistent or transient, based on whether a source was permanently visible by a given instrument or only occasionally. However, such a subdivision is an oversimplification, as sources have been found to vary over large ranges of timescales and amplitudes \citep{levine2011}. One type of variability, first observed in dwarf novae, is attributed to a thermal-viscous instability in the accretion disk \citep{osaki1974,smak1984,lasota2001}. As the disk grows in mass and heats up to close to the ionization temperature of the dominant element, the opacity strongly increases and radiation is trapped, and the disk's surface density approaches a threshold value. The temperature continues to rise more rapidly and the disk enters the hot state. The resulting high viscosity enhances the outward angular momentum transport in the disk, allowing for a higher accretion rate. This can be observed as an outburst, after which the cycle repeats. For UCXBs the mass transfer rate (and thus the disk temperature) is expected to decrease as a function of orbital period, suggesting that systems with orbital periods above $\sim \! 30$ min should undergo this instability in their outer disks \citep[see][]{deloye2003}.

\subsection{Present research}

We investigate the bolometric luminosity distributions of the known UCXB population including candidates with tentative orbital periods, which show the fraction of time a source spends at a given luminosity. The relation between the luminosity distribution and the average luminosity may give insight in the way transient and persistent behavior should be interpreted.

We do not investigate periodicities in the light curves since this has been done already \citep{levine2011}, and periodicities are not relevant for the present purposes.

\section{Method}
\label{method}

Here we describe the known UCXB population and their observational records, and how we use these data.

\subsection{Known UCXB population}

UCXBs are very rare objects. Presently $13$ have been identified with high confidence (by finding the orbital period), all Galactic. These are located in globular clusters, the Galactic Bulge and the Galactic Plane. Table \ref{table:params} presents an updated version of the UCXB and UCXB candidate overviews in \citet{zand2007}, \citet{lasotaetal2008} and \citet{nelemans2010b}.

\begin{table*}
\caption{Known ultracompact X-ray binaries and candidates. The four groups separated by horizontal lines are confirmed UCXBs, UCXB candidates with tentative orbital period, UCXB candidates with a low ratio of optical to X-ray luminosity and UCXB candidates with persistent behavior at low luminosity, respectively. The columns list source name, orbital period $P_\mathrm{orb}$, distance $d$, location within the Galaxy, chemical composition and indication of accretor type (`msp' = millisecond pulsar, `burst' = Type 1 X-ray burst). Sources with coordinates very near the Galactic Bulge are assumed to be located in the Bulge. Distances are taken from \citet{liu2007} unless indicated otherwise. In the references column, `p' refers to orbital period, `c' to composition, `d' to distance and `a' to accretor signal.}
\label{table:params}
\centering
\begin{tabular}{lllllllll}
\hline\hline
Name   & $P_\mathrm{orb}\ (\mbox{min})$     & $d$ (kpc)         & Location  & Composition           & Accretor & References \\
\hline
\object{4U 1820--30}            & $11.42$   & $7.6 \pm 0.4$     & \object{NGC 6624}  & He (burst-model) & burst & 1-p, 2-d, 3,4,5-c, 6-a \\
\object{4U 0513--40}            & $17$      & $12.2$            & \object{NGC 1851}  & He           & burst  & 7-pc, 8-a\\
\object{2S 0918--549}           & $17.4$    & $4.8 \pm 0.6$     & Disk      & C,O? (opt), He (burst) & burst & 9-p, 10,11-d, 12,13-c, 14-a\\
\object{4U 1543--624}           & $18.2$    & $\sim \! 7?$      & Disk      & C,O? (opt), O (X)     & --     & 15-pd, 12,16-c\\
\object{4U 1850--087}           & $20.6$    & $8.2 \pm 0.6$     & \object{NGC 6712}  & Ne-excess    & burst  & 17-p, 2-d, 18-c, 19,20-a\\
\object{M 15 X-2}               & $22.58$   & $10.4$            & \object{M 15}      & He,C (UV)    & burst  & 21-pc, 22-d, 23,24-a\\ 
\object{XTE J1807--294}         & $40.07$   & $8.3 \pm 1.5$?    & Bulge?    &                       & msp    & 25-p, 26-a \\
\object{4U 1626--67}            & $41.4$    & $\sim \! 8$       &           & C,O (opt), O (UV,X)   & pulsar & 27-pa, 28-d, 29,30,31,32-c \\
\object{XTE J1751--305}         & $42.42$   & $8.3 \pm 1.5$?    & Bulge?    &                       & msp    & 33-pa \\
\object{XTE J0929--314}         & $43.58$   & $10 \pm 5$        &           & He,C,N                & msp    & 34-pca, 35,36-d, 29-c \\
\object{4U 1916--05}            & $49.48$   & $8.9 \pm 1.3$     &           & He,N (opt)            & burst  & 37,38-p, 39-d, 29-c, 40,41-a\\
\object{SWIFT J1756.9--2508}    & $54.70$   & $8.3 \pm 1.5$?    & Bulge?    & He (model)            & msp    & 42-pca \\
\object{NGC 6440 X-2}           & $57.3$    & $8.5 \pm 0.4$     & \object{NGC 6440}  &              & msp    & 43-pa, 44-d \\ 
\hline
\object{4U 1728--34} (GX 354--0) & $10.77$?  & $5.3 \pm 0.8$     &           &                       & burst,msp & 45,46-p, 11-d, 47,48,49-a \\
\object{NGC 6652 B}             & $43.6$?   & $10.0$            & \object{NGC 6652} &               & burst  & 50-p, 22-d, 51-a \\ 
\object{4U 0614+091}            & $51.3$?   & $3.2$             &           & C,O (opt), O (X)      & burst,msp  & 52,53,54-p,55-d,12,30,56-c,57,58,59,60-a \\
\hline
\object{1A 1246--588} & & $4.3$ & & & & 61,62-d \\
\object{4U 1812--12} & & & & & & 61 \\
\object{4U 1822--000} & & & & & & 63 \\
\object{4U 1905+000} & & 8 & & & & 64 \\
\object{$\omega\,$Cen qLMXB} & & & \object{NGC 5139} & & & 65 \\
\hline
\object{1RXS J170854.4-321857} & & & & & & 66 \\ 
\object{SAX J1712.6-3739} & & & & & & 63 \\
\object{1RXS J171824.2-402934} & & & & & & 66 \\
\object{4U 1722-30} & & & \object{Terzan 2} & & & 63 \\
\object{1RXS J172525.2-325717} & & & & & & 63 \\
\object{SLX 1735-269} & & & & & & 63 \\
\object{SLX 1737-282} & & & & & & 63 \\
\object{SLX 1744-299} & & & & & & 63 \\
\hline
\end{tabular}
\tablebib{
1 \citet{stella1987};
2 \citet{kuulkers2003};
3 \citet{bildsten1995};
4 \citet{podsiadlowski2002}; 
5 \citet{cumming2003}; 
6 \citet{grindlay1976};
7 \citet{zurek2009};
8 \citet{forman1976};
9 \citet{zhong2011};
10 \citet{cornelisse2002}; 
11 \citet{jonker2004}; 
12 \citet{nelemans2004}; 
13 \citet{zand2005}; 
14 \citet{jonker2001};
15 \citet{wang2004};
16 \citet{madej2011}; 
17 \citet{homer1996};
18 \citet{juett2001}; 
19 \citet{swank1976};
20 \citet{hoffman1980};
21 \citet{dieball2005};
22 \citet[][2010 edition]{harris1996};
23 \citet{dotani1990};
24 \citet{vanparadijs1990};
25 \citet{markwardt2003b};
26 \citet{markwardt2003c};
27 \citet{middleditch1981};
28 \citet{chakrabarty1998b};
29 \citet{nelemans2006}; 
30 \citet{werner2006}; 
31 \citet{schulz2001}; 
32 \citet{homer2002}; 
33 \citet{markwardt2002};
34 \citet{galloway2002};
35 \citet{wijnands2005};
36 \citet{liu2007};
37 \citet{white1982};
38 \citet{walter1982};
39 \citet{galloway2008};
40 \citet{becker1977};
41 \citet{lewin1977};
42 \citet{krimm2007};
43 \citet{altamirano2010};
44 \citet{ortolani1994};
45 \citet{galloway2010};
46 \citet{wilsonhodge2012arXiv};
47 \citet{lewin1976};
48 \citet{hoffman1976};
49 \citet{strohmayer1996};
50 \citet{deutsch2000};
51 \citet{zand1998b};
52 \citet{shahbaz2008};
53 \citet{hakala2011};
54 \citet{zhang2012};
55 \citet{kuulkers2010};
56 \citet{madej2010};
57 \citet{swank1978};
58 \citet{brandt1992};
59 \citet{ford1997}; 
60 \citet{strohmayer2008}; 
61 \citet{bassa2006};
62 \citet{zand2008};
63 \citet{zand2007}; 
64 \citet{jonker2006};
65 \citet{haggard2004};
66 \citet{zand2005a}.
}
\end{table*}

\subsection{Observations}
\label{observ}

We use observations by the \emph{Rossi X-ray Timing Explorer} All-Sky Monitor (\emph{RXTE} ASM) \citep{bradt1993,levine1996}. The ASM has collected $16$ yr X-ray light curves and therefore gives insight in the source behavior over a large range of timescales. By its very nature of being an all-sky monitor, the ASM is an unbiased sampler of light curves, i.e., its observations are not driven by source behavior.
Pointed observations by e.g. \emph{Chandra} or \emph{RXTE} PCA have a higher sensitivity than the ASM, but are available far less regularly and often during atypical source behavior. Hence, these observations cannot be used to gain insight in the `normal' behavior over timescales of years.

\subsubsection{\emph{RXTE} All-Sky Monitor}
\label{method_asm}

The \emph{RXTE} ASM has monitored the sky in the $2 - 10\ \mbox{keV}$ energy range from January 5, 1996 to January 3, 2012 (MJD $50087-55929$). Individual exposures, called \emph{dwells}, have an integration time of typically $90$ s. The intervals between dwells are irregular and on average $\sim \! 2$ hr. Light curves have been published in the ASM Products Database\footnote{\url{http://heasarc.gsfc.nasa.gov/docs/xte/asm_products.html}} for many resolved X-ray sources, including $12$ of the $13$ confirmed UCXBs.
X-ray photon count rates in the ASM energy range have been converted to bolometric flux using the conversion of $1\ \mbox{ASM counts s}^{-1} = 7.7 \times 10^{-10}\ \mbox{erg cm}^{-2}\ \mbox{s}^{-1}$ by \citet{zand2007} (with an error of a factor of $\sim \! 2$) and to bolometric luminosities using the distance estimates in Table \ref{table:params}. The ASM does not provide a light curve for \object{NGC 6440 X-2} \citep{heinke2010} and UCXB candidate \object{NGC 6652 B}. Both share their host globular cluster with a brighter X-ray source. The latter, \object{NGC 6652 B}, is too close to \object{XB 1832--330} to be resolved by the ASM \citep{heinke2001}. Furthermore, \object{M 15 X-2} is close to another source, \object{AC 211}, in the globular cluster \object{M 15}. The ASM does not resolve these two sources, but because \object{AC 211} is $2.5$ times fainter than \object{M 15 X-2} \citep{dieball2005} we still use the combined X-ray data (known as \object{4U 2127+119}) in this paper.

\subsection{Data analysis}

In order to calculate the luminosity distribution of a source, we use two methods, each based on individual dwells. The first method selects only those dwells that have a signal-to-noise ratio ($S/N$) above a fixed threshold, while the second method uses time-averaged luminosity to estimate the faint-end variability.

A commonly used technique is grouping dwells into time bins of a fixed duration to statistically improve the $S/N$, thereby revealing fainter behavior. We do not apply this because in order to benefit from this, the bin duration should be on the order of days, and variability on shorter timescales will be lost. In particular the bright-end behavior will be affected as high count rate dwells are combined with faint or noisy dwells.

\subsubsection{Signal-to-noise threshold}
\label{methodsnr}

The bright-end part of the luminosity distribution is best revealed by using the shortest available observations for maximum sensitivity, i.e., individual dwells. The downside of this is that many individual dwells have a count rate near the noise level.

We use a strict criterion to select statistically significant dwells. For a given source, only dwells with $S/N > \sigma$ are selected, where $\sigma$ is found by assuming that a fraction $1/n$ of the $n$ dwells lies outside $\sigma$ times the standard deviation of a Gaussian probability distribution \citep{forbes2010book}, i.e.,

\begin{equation}
    \label{outlier}
    1 - \mathrm{erf}\left(\frac{\sigma}{\sqrt{2}}\right) = \frac{2}{n}.
\end{equation}
The factor $2$ on the right-hand side appears due to asymmetry; only positive rates are considered. Dwells failing this criterion are discarded. The critical $S/N \approx 4.2$ for $n = 7 \times 10^{4}$, a typical ASM value. The remaining sample is statistically expected to contain $1$ data point that is not `real' (in the sense that a distribution fully dominated by noise would be expected to have $1$ outlier in its bright-end tail sufficiently far from the mean). This is the advantage of our method; the selected dataset consists of (almost) exclusively `real' detections and therefore it can be used to find a lower limit of the system's average bolometric luminosity.
However, this dataset is not complete; at the times when the detection has an insufficient $S/N$ to pass the criterion, the actual luminosity of the source is unlikely to be zero. This luminosity can be estimated by an extrapolation of the known part of the distribution \citep[][chap.~3]{press1992book}. We adopt the simple approach of extrapolating the bright-end part towards lower luminosities by a power-law function, as suggested by the relatively straight curves of the luminosity distribution when plotted as a log-log graph (see also Sect. \ref{faint}). The interval over which we calculate the slope ranges from a fraction that is a factor $10^{0.2}$ below the fraction of time during which the source is detected significantly, to a fraction of $10^{-4.5}$. The faint-end and bright-end parts are excluded for practical reasons that will be explained in Sect. \ref{resultssnr}. From the extrapolation, we can integrate the (long-term) average luminosity. The validity of this extrapolation can be verified by comparing the resulting time-averaged luminosity with the time-averaged ASM luminosity and theoretical models.

\begin{figure}
\resizebox{\hsize}{!}{\includegraphics{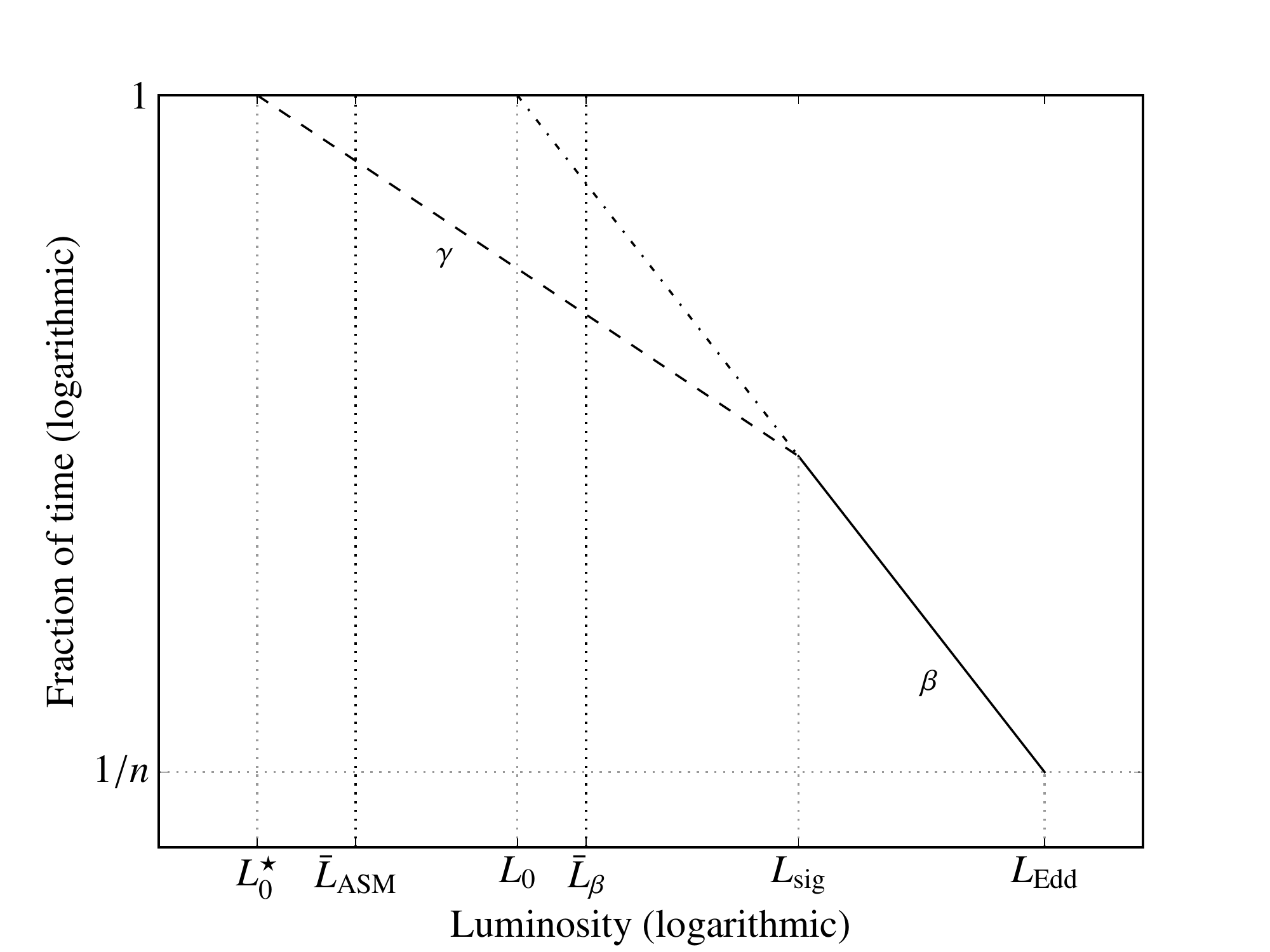}} 
\caption{Sketch of a cumulative luminosity distribution. The highly significant bright-end distribution (solid) and its linear extrapolation (dash-dotted) have slope $\beta$ and are discussed in Sect. \ref{methodsnr}. The extension of the bright-end distribution with slope $\gamma$ (dashed) is discussed in Sect. \ref{methodtwo}. The variables on the axes are defined in these two sections. In general the $\gamma$-slope can be either flatter than (shown), equal to, or steeper than the $\beta$-slope.}
\label{fig:slope}
\end{figure}

If the logarithmic slope of a cumulative luminosity distribution (i.e., a time-above-luminosity against luminosity curve)\footnote{Not to be confused with the X-ray luminosity function of a population of X-ray binaries.} is $\beta$, then the luminosity distribution (i.e., time-at-luminosity against luminosity curve) function slope is $\beta - 1$. Combined with the minimum luminosity $L_{0}$, which is the luminosity $L$ where the cumulative distribution has a fraction $1$, and the Eddington luminosity $L_\mathrm{Edd}$ (see Fig. \ref{fig:slope}), this yields the time-averaged bolometric luminosity ($\beta < -1$) \citep[][chap.~3]{bremaud1997book}

\begin{equation}
    \label{avglum}
    \bar{L}_{\beta} = \frac{\int_{L_{0}}^{L_\mathrm{Edd}} L \times L^{\beta-1} \mathrm{d}L}{\int_{L_{0}}^{L_\mathrm{Edd}} L^{\beta-1} \mathrm{d}L} = \frac{\beta}{\beta + 1} \left( \frac{L_\mathrm{Edd}^{\beta + 1} - L_{0}^{\beta + 1}}{L_\mathrm{Edd}^{\beta} - L_{0}^{\beta}} \right) \approx \frac{\beta}{\beta + 1} L_{0}.
\end{equation}
The approximation holds if $(L_\mathrm{Edd}/L_{0})^{\, \beta + 1} \ll 1$ (which is the case for all known UCXBs except \object{2S 0918--549}).

\subsubsection{Faint-end power-law function}
\label{methodtwo}

Instead of assuming that the source luminosity behavior can be described by a single power-law function as we did in Sect. \ref{methodsnr}, here we assume that the source behavior at the times that the source is fainter than $L_\mathrm{sig}$ (defined as the faint-end boundary of the range over which $\beta$ is calculated) can be described by a second power-law function, with slope $\gamma \le 0$, which in general is different from the power-law slope $\beta$ between $L_\mathrm{sig}$ and $L_\mathrm{Edd}$ (see Fig. \ref{fig:slope}). The two power-law functions meet in $L_\mathrm{sig}$. We solve for $\gamma$ and the corresponding value of $L_{0}^{\star}$ (the luminosity where the cumulative distribution based on both slopes $\beta$ and $\gamma$ has a fraction $1$) by assuming that the integrated luminosity between $L_{0}^{\star}$ and $L_\mathrm{Edd}$ equals the time-averaged ASM bolometric luminosity $\bar{L}_\mathrm{ASM}$ (using all dwells),

\begin{equation}
    \label{twoslopes}
    \frac{\int_{L_{0}^{\star}}^{L_\mathrm{Edd}} L \times f(L)\, \mathrm{d}L}{\int_{L_{0}^{\star}}^{L_\mathrm{Edd}} f(L)\, \mathrm{d}L} = \bar{L}_\mathrm{ASM},
\end{equation}
where the luminosity distribution function

\begin{equation}
    \begin{array}{r}
        \label{twoslopes2}
        f(L) =
        \left\{ \begin{array}{ll}
        L^{\gamma-1}                                    & \mbox{ if \ } L_{0}^{\star}  \le L \le L_\mathrm{sig}, \\
        L_\mathrm{sig}^{\gamma - \beta} L^{\beta-1}     & \mbox{ if \ } L_\mathrm{sig} \le L \le L_\mathrm{Edd}.
        \end{array} \right.
    \end{array}
\end{equation}
We used the continuity of $f$ in $L_\mathrm{sig}$ to find the factor $L_\mathrm{sig}^{\gamma - \beta}$ in front of $L^{\beta - 1}$.\footnote{It follows that also the cumulative luminosity distribution is continuous in $L_\mathrm{sig}$.}

A steep $\gamma$-slope (very negative $\gamma$) means that the source has a relatively constant luminosity when it is fainter than $L_\mathrm{sig}$, and spends much time at luminosities not far below $L_\mathrm{sig}$. A flat $\gamma$-slope on the other hand points to a generally faint source that spends a considerable amount of time far below $L_\mathrm{sig}$, and is highly variable.

This method does not work for two sources, \object{XTE J1807--294} and \object{XTE J0929--314}, because $\bar{L}_\mathrm{ASM} < 0$ for those.\footnote{This is due to a systematic error. Faint sources, or sources far outside the center of the field of view, have an increased probability to suffer from an imperfect solution for the brightness and position of the sources in the ASM field of view, and can end up with a consistently negative count rate, resulting in a negative average count rate over all dwells \citep{levine1996}.} Hence, the contribution to $\bar{L}_\mathrm{ASM}$ from luminosities below $L_\mathrm{sig}$ would have to be negative. In reality this suggests that these sources must be very faint or off when not detected significantly, but more precise conclusions are not possible.

\subsection{Theoretical evolution}
\label{theory}

Mass transfer in UCXBs is driven by angular momentum loss via gravitational wave radiation, since angular momentum loss via magnetic braking is much weaker at such a short orbital period \citep{landau1975,rappaport1983}. Therefore the mass transfer rate strongly depends on the donor mass and orbital separation. During most of the evolution, the donor is degenerate and expands with mass loss. Since the donor keeps filling its Roche lobe and the ratio between donor and accretor mass decreases, the orbital separation expands with time. The decreasing donor mass and increasing orbital separation cause the gravitational wave emission and the corresponding mass transfer rate to decrease with time.

A fit of the mass transfer rate $\dot{M}$ (defined $\ge 0$) in an UCXB with a zero-temperature helium white dwarf donor and an $1.4\ M_{\odot}$ neutron star accretor based on theoretical evolutionary tracks \citep{vanhaaften2012} and optimized for an orbital period $P_\mathrm{orb} \le 1$ hr is

\begin{equation}
    \label{mdot}
    \dot{M} = 1.2 \times 10^{-12} \left(\frac{P_\mathrm{orb}}{\mathrm{hr}}\right)^{-5.2} M_{\odot}\ \mathrm{yr}^{-1}.
\end{equation}
A simple estimate of the time-averaged bolometric luminosity $\bar{L}$ is

\begin{equation}
    \label{lum}
    \bar{L} = \frac{G M_\mathrm{a} \dot{M}}{2 R_\mathrm{a}}
\end{equation}
where $G$ is the gravitational constant and $M_\mathrm{a}$ and $R_\mathrm{a}$ the accretor mass and radius, respectively. For $M_\mathrm{a} = 1.4\ M_{\odot}$ and $R_\mathrm{a} = 12$ km Eqs. (\ref{mdot}) and (\ref{lum}) combine to

\begin{equation}
    \label{lum2}
    \bar{L} = 6 \times 10^{33} \left(\frac{P_\mathrm{orb}}{\mathrm{hr}}\right)^{-5.2} \mathrm{erg\ s^{-1}},
\end{equation}
which is shown in Sect. \ref{res_avglum}.

Young UCXBs can contain a helium burning donor \citep{savonije1986}, in which case the mass transfer rate at a given orbital period is much higher than in the case of a white dwarf donor, due to the higher donor mass. During this evolutionary stage the orbit shrinks because the donor shrinks sufficiently rapidly as it loses mass. Once the donor has lost enough mass, core fusion is extinguished and a hot white dwarf donor remains. The system goes through a period minimum. Further mass loss and cooling cause the evolution to approach the zero-temperature white dwarf donor evolution.

UCXBs may also form via an evolved main sequence donor scenario \citep{podsiadlowski2002}, but only for highly fine-tuned initial conditions \citep{sluys2005a}, therefore we do not consider this scenario.

\section{Results}
\label{results}

\subsection{Signal-to-noise threshold}
\label{resultssnr}

\begin{figure}
\resizebox{\hsize}{!}{\includegraphics{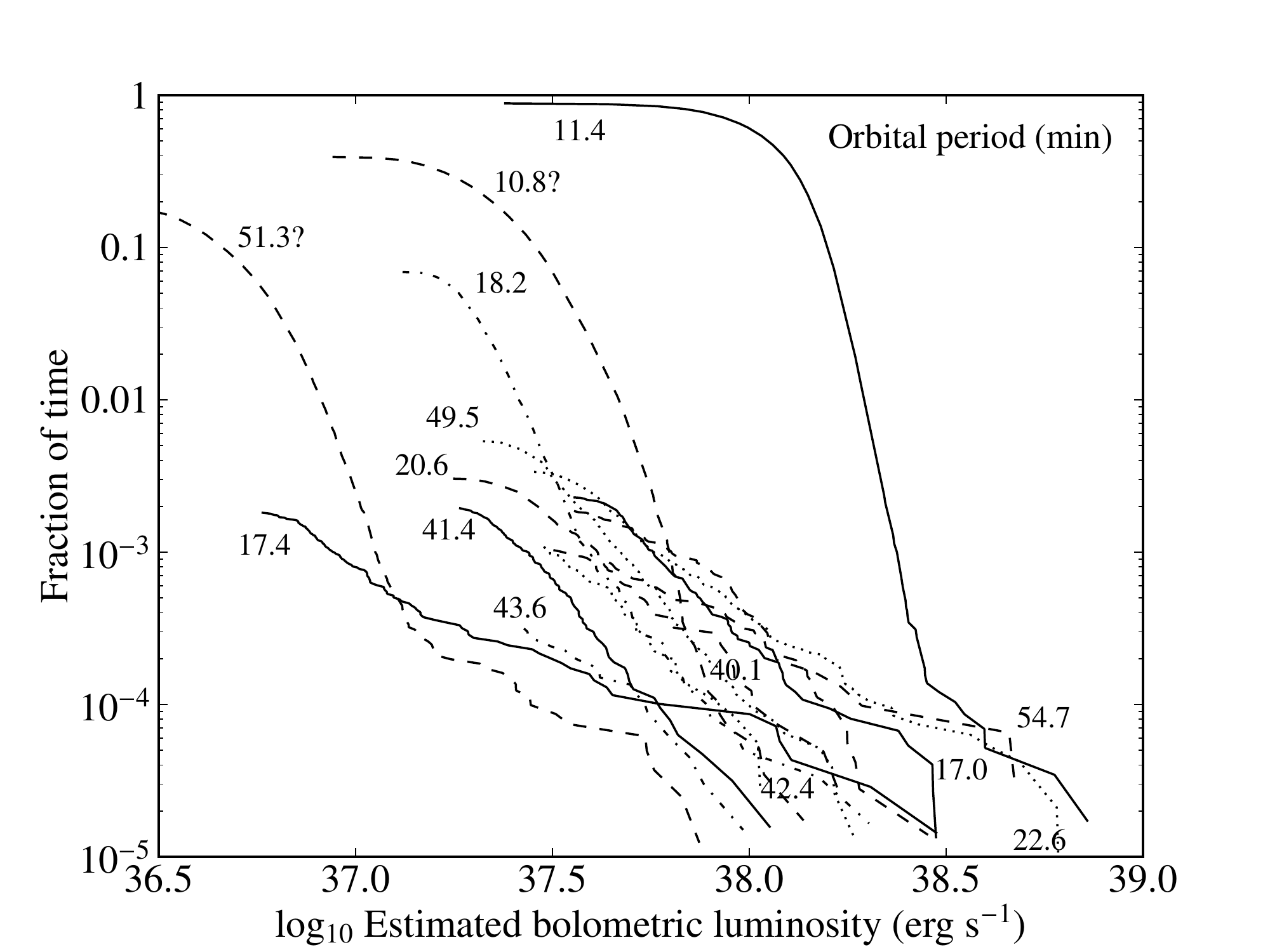}} 
\caption{Cumulative probability distribution of UCXB bolometric luminosity based on ASM data (fraction of time that a source emits above a given luminosity) for $14$ UCXBs including two candidates with a tentative orbital period. Only count rates that are almost certainly real (their significant detection is almost certainly not the result of noise) have been used (Sects. \ref{methodsnr} and \ref{resultssnr}). The labels indicate the orbital period in min of the corresponding systems. Different line styles are used only to distinguish different systems.}
\label{fig:signi}
\end{figure}

By selecting only dwells that contain a highly significant source detection as described in Sect. \ref{methodsnr}, a sensitive bright-end cumulative luminosity distribution can be constructed, as shown in Fig. \ref{fig:signi}. This figure shows how often a source emits above a given luminosity. Strictly speaking it shows the fraction of dwells, but this is similar to the fraction of time because of their unbiased timing. Here, the total number of dwells includes those that were discarded.

From the vertical locations of the faint-end points of the curves it can be seen that typically only $\sim \! 0.1 - 1\%$ of the dwells pass the high-threshold criterion, emphasizing the fact that only the brightest phases of a source can be studied using this method.
The short orbital period systems are generally visible more often than the long (in the context of ultracompact binaries) orbital period systems, as expected from theory. \object{4U 1820--30} (orbital period $11.4$ min) is the only source that is clearly observed nearly all the time (at $88\%$ of the time).
Most UCXBs are strongly variable, in the sense that curves are far from vertical. The most extreme case is \object{2S 0918--549} ($17.4$ min), which has a very flat logarithmic slope of $\beta \approx -1$ in Fig. \ref{fig:signi}. Most slopes, however, are steeper with $\beta \lesssim -2$ (see Table \ref{table:slopes}).

Since luminosity is used on the horizontal axis rather than flux or count rate, and given that the faintest data points of the cumulative distributions have a roughly similar count rate (determined by the instrument's detection limit), the source distance determines at which luminosity the faint end of each curve is located. The UCXB candidate \object{4U 0614+091} is by far the nearest system at an estimated distance of $3.2$ kpc \citep{kuulkers2010}, therefore this source can be observed down to a lower intrinsic luminosity than the more distant sources.

An artifact of the technique used in Fig. \ref{fig:signi} can be seen at the faint ends of several curves, where at low luminosities the slope flattens towards a fraction of time usually much lower than $1$. This behavior most likely does not have a physical origin and happens because at these faint luminosities, by chance a small number of data points will have an unusually low error and therefore pass the threshold, while the vast majority of data points with approximately the same count rate fails the threshold. This extends the curves further to low luminosities almost horizontally. This effect is more pronounced for sources that have more significant data points (corresponding to a fraction of time $\gtrsim \! 10^{-3}$), as those have a higher probability to have outliers (detections with an unusually low error).

Even though the observation schedule of the ASM is predetermined, irregular observation intervals may introduce an extra bias when an above- or below-average observation frequency coincides with a period of abnormal source behavior. Since Fig. \ref{fig:signi} uses individual dwells regardless of their timing, the shape and position of these distributions could potentially be different from their shape and position in the case of using bins of fixed duration. This effect appears to be weak since there is almost no correlation between interval duration and count rate. At a fraction below $\sim \! 10^{-4}$ the slopes of the curves become less reliable due to the low number of dwells the curves are based on, in particular there is the possibility of atypical interval lengths surrounding the highest count rates. Even in the case of regular sampling there would be some bias at the highest count rates, because even then small number statistics allow a source to behave differently during observations compared to in between observations.

Although some luminosities in Fig. \ref{fig:signi} appear to go above the Eddington limit for a neutron star, we note that given the uncertainties in both the distance and the ASM count to flux conversion, these luminosities are consistent with being sub-Eddington within errors.

\subsubsection{Faint-end behavior}
\label{faint}

The nearly constant slopes over a large part of the curves in Fig. \ref{fig:signi} (excluding the faint-end flattened part and the statistically uncertain bright-end part of these curves for the reasons discussed in Sect. \ref{resultssnr}) suggest the possibility that this behavior may continue into the faint-end phases of the sources that have not been detected reliably. By extrapolating the middle parts of these curves with a power law, the time-averaged luminosity can be expressed in essentially only the slope and the minimum luminosity, unless the slope is close to $\beta = -1$ (Sect. \ref{methodsnr}). In Sect. \ref{res_avglum} the result will be compared with the time-averaged luminosity and with theory.

\subsection{Average luminosity}
\label{res_avglum}

\begin{figure}
\resizebox{\hsize}{!}{\includegraphics{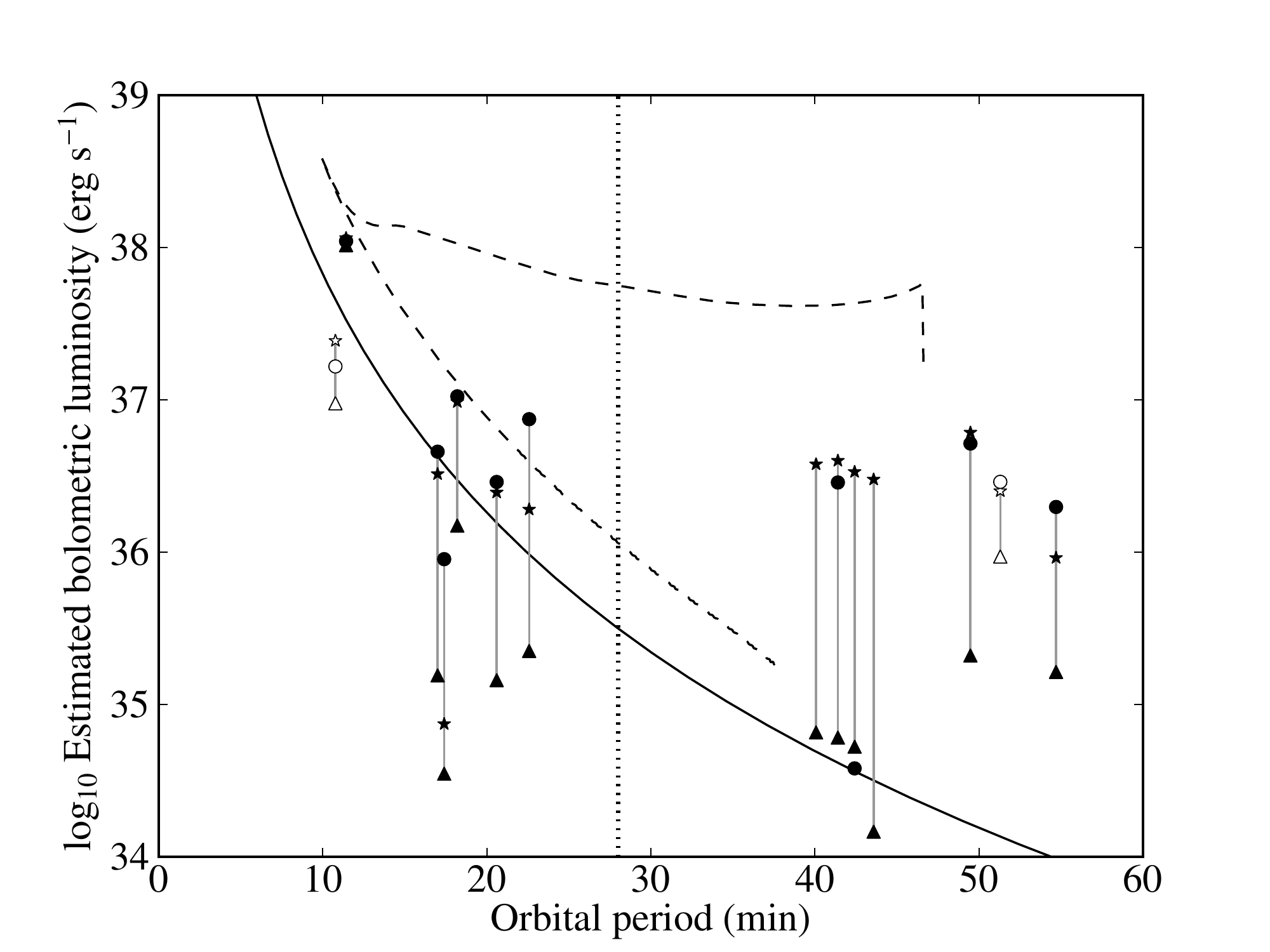}} 
\caption{UCXB time-averaged bolometric luminosity against orbital period. The solid curve represents the modeled evolution of an UCXB with a zero-temperature helium white dwarf donor (Eq. \ref{lum2}). The dashed curve represents the modeled evolution of an initially detached system with an $0.65\ M_{\odot}$ helium burning donor starting at an orbital period of $100$ min, which starts mass transfer via Roche-lobe overflow at an orbital period of $46$ min \citep{yungelson2008}. Both systems have an $1.4\ M_{\odot}$ neutron star accretor. The triangles represent the absolute lower bounds on the luminosity based on only the highly significant ASM dwells and assuming zero luminosity at all other times (Sects. \ref{methodsnr} and \ref{resultssnr}, Fig. \ref{fig:signi}). The stars above these represent the power-law extrapolation of the bright-end part of the cumulative luminosity distribution $\bar{L}_\mathrm{\beta}$ (Sect. \ref{faint}). The circles represent the time-averaged bolometric luminosities based on all dwells $\bar{L}_\mathrm{ASM}$. Filled symbols correspond to confirmed UCXBs, open symbols correspond to UCXB candidates with a tentative orbital period. Symbols at the same orbital period correspond to the same source and are connected by a gray line for clarity. For two sources the average count rate is negative so the circle is not shown. The vertical dotted line at $28$ min represents the orbital period above which a helium accretion disk becomes subject to the thermal-viscous instability, based on the zero-temperature white dwarf donor model by \citet{vanhaaften2012}.}
\label{fig:avglum}
\end{figure}

The variability behavior of the UCXBs presented in Sect. \ref{resultssnr} can be summarized and compared by using the time-averaged luminosity. Figure \ref{fig:avglum} shows several bolometric luminosity estimates for the known UCXBs. The average luminosity over all dwells (circles) can be seen as the best estimate of these three taking all uncertainties into account, unless, of course, when the average count rate is negative. In general, the luminosity estimates based on power-law extrapolation (Sect. \ref{faint}) and the time-averaged value over all dwells are of the same order of magnitude.

Figure \ref{fig:avglum} also compares the average-luminosity estimates with theoretical UCXB tracks, with both a zero-temperature white dwarf donor and a helium burning donor. The helium burning donor system has typical initial conditions among the ranges simulated by \citet{yungelson2008}.

The average luminosity of the short-period UCXBs (below $\sim \! 30$ min) agrees well with the theoretical model based on the mass transfer rate (Of these, \object{M 15 X-2} is contaminated by a fainter source, see Sect. \ref{method_asm}). Based on the ASM data, there is no reason to suspect variability on timescales longer than $16$ yr in these sources. The extrapolated power-law estimate closely matches both the average luminosity and the theoretical prediction. \object{4U 1820--30} ($11.4$ min) is the only source with a very well-determined average luminosity, which fits well with the representative helium burning donor track. As this source may have a negative period derivative \citep{tan1991}, it could correspond to the upper branch of a helium donor track, but this is uncertain. \object{2S 0918--549} ($17.4$ min), which has the flattest slope in Fig. \ref{fig:signi}, has a correspondingly low luminosity estimate based on the extrapolated power law, much lower than even the zero-temperature white dwarf donor model. This means that either the source has gone through an unusually faint period of time the last $16$ years, or the extrapolation underestimates its faint-end luminosity.

The vertical dotted line in Fig. \ref{fig:avglum} represents the critical orbital period for thermal-viscous disk instability, in the case of a \citet{dunkel2006} helium accretion disk and a zero-temperature white dwarf donor. It lies in a large apparent period gap ranging from $23$ to $40$ min.

Above this gap, the four UCXBs with orbital periods between $40$ and $45$ min have lower bounds (triangles) close to the zero-temperature donor track. This is consistent with them being (very) faint when not detected, i.e., transient, as is expected for systems with these orbital periods. The power-law extrapolations (stars), which give a much higher luminosity estimate, would not apply then.

However, the three UCXBs, including one candidate, with an orbital period longer than $\sim \! 50$ min all have a lower bound on their luminosity that exceeds both the zero-temperature white dwarf track and the (extrapolated) degenerate part of the helium donor track. Different initial conditions can result in a slightly higher helium donor track \citep{yungelson2008} so some of these systems can still be consistent with having a (partially) degenerate donor. One source, \object{4U 0614+091}, cannot be explained by having a semi-degenerate donor as its lower bound stands out above the degenerate track too much. \object{4U 0614+091} has an uncertain orbital period, but if this period is correct then this suggests the system has a helium burning donor.
Another possibility is that this source was exceptionally bright during the past $16$ yr, and that its average luminosity over a longer time is lower. Similarly, the time-averaged ASM luminosities of all three UCXBs with an orbital period above $\sim \! 50$ min, as well as that of \object{XTE J1751--305} at $42$ min indicate that these systems are much brighter than predicted by the theory for UCXBs with a degenerate donor, but still fainter than expected for UCXBs with a helium burning donor. In several cases the time-averaged luminosity is nearly equal to the power-law extrapolation estimate, similar to the short-period systems. It is unlikely that all three known systems with orbital period $\gtrsim \! 50$ min are in a bright state by chance. An explanation could be that the theory described in Sect. \ref{theory} underestimates the time-averaged mass transfer rate of long-period systems. Alternatively, UCXBs at long periods may be variable on a timescale much longer than $16$ yr, so that during the ASM observations some systems were much brighter and many others (much) fainter than the average luminosity. Such a selection effect would lead us to observe only the atypically bright systems.

\subsection{Faint-end power-law function}
\label{resultstwo}

If the extrapolated power-law method $\bar{L}_\mathrm{\beta}$ (Sect. \ref{methodsnr}, stars in Fig. \ref{fig:avglum}) underestimates the time-averaged luminosity $\bar{L}_\mathrm{ASM}$\footnote{The time-averaged luminosity is calculated based on the average count rate over all individual dwells (regardless of their timing), which is a good approximation given the unbiased timing and large datasets.} (circles in Fig. \ref{fig:avglum}), then the slope of the cumulative luminosity distribution is expected to steepen below $L_\mathrm{sig}$, i.e., $\gamma < \beta$. This means that the source is brighter and more constant at low luminosities (fainter than the significant detections) than suggested by extrapolating the behavior during the bright phases. Conversely, if the extrapolated power-law method overestimates the time-averaged luminosity then $\gamma > \beta$ and the slope flattens below $L_\mathrm{sig}$. In this case the source is more variable when faint than when bright. Table \ref{table:slopes} shows the values of $\beta$ for the known UCXBs as well as $\gamma$ for the sources with a sufficiently high (in practice, positive) time-averaged luminosity. Steepening of the faint-end power-law slope happens for all short-period sources (below $\sim \! 30$ min) except \object{4U 1820--30} at $11.4$ min. The latter may be caused by the fact that this source is detected significantly a very high fraction ($88\%$) of the time. On the other hand, almost all long-period sources (above $\sim \! 30$ min) behave the opposite way, with a flattened faint-end slope. The one certain exception is \object{SWIFT J1756.9--2508} at $54.7$ min, while \object{4U 0614+091} would be a second exception if its orbital period would be confirmed.

\begin{table}
\caption{Power-law slopes $\beta$ and $\gamma$ describing the cumulative luminosity distribution (see Sects. \ref{methodsnr} and \ref{methodtwo}) for the known UCXB population based on the \emph{RXTE} ASM data. The two systems below the line are UCXB candidates, with an uncertain orbital period. If no $\gamma$-value is given, the source is too faint on average for this method to work.}
\label{table:slopes}
\centering
\begin{tabular}{llcc}
\hline\hline
Name      & $P_\mathrm{orb}\ (\mbox{min})$      & $\beta$      & $\gamma$ \\
\hline
\object{4U 1820--30}            & $11.42$       & $-7.4$       & $-3.7$ \\ 
\object{4U 0513--40}            & $17$          & $-2.0$       & $-2.3$ \\%
\object{2S 0918--549}           & $17.4$        & $-1.03$      & $-2.5$ \\%
\object{4U 1543--624}           & $18.2$        & $-3.8$       & $-4.3$ \\%
\object{4U 1850--087}           & $20.6$        & $-1.9$       & $-2.1$ \\%
\object{M 15 X-2}               & $22.58$       & $-1.6$       & $-3.0$ \\
\object{XTE J1807--294}         & $40.07$       & $-2.4$       & N/A \\
\object{4U 1626--67}            & $41.4$        & $-3.0$       & $-2.5$ \\
\object{XTE J1751--305}         & $42.42$       & $-2.5$       & $-0.55$ \\
\object{XTE J0929--314}         & $43.58$       & $-2.8$       & N/A \\
\object{4U 1916--05}            & $49.48$       & $-2.8$       & $-2.5$ \\
\object{SWIFT J1756.9--2508}    & $54.70$       & $-1.3$       & $-1.7$ \\%
\hline
\object{4U 1728--34}            & $10.77$?      & $-6.3$       & $-0.66$ \\ 
\object{4U 0614+091}            & $51.3$?       & $-3.2$       & $-4.1$ \\%
\hline
\end{tabular}
\end{table}

\section{Implications for population studies}

The results concerning the luminosity behavior presented in Sect. \ref{results} can be used in modeling a population of UCXBs. Based on its evolutionary state, a modeled UCXB can be associated with a real UCXB, and its luminosity and variability can be estimated empirically by drawing from the observational luminosity distribution, without need to understand the theory behind the variability. This method rests on the assumption that the available observational record is typical for such systems, which may not be the case for the long-period UCXBs observed by the ASM, as shown in Fig. \ref{fig:avglum} and noted in Sect. \ref{res_avglum}.

Since many UCXBs are only visible a fraction of the time above a given luminosity as shown by Fig. \ref{fig:signi}, at any given time we will only see a fraction of the UCXB population, using an instrument with a given sensitivity. If the instrument is the \emph{RXTE} ASM itself, only a small fraction of the population can be seen at any given time. For a more sensitive instrument, the shape of the luminosity distribution determines the characteristics of the visible population. A flat slope at the faint end of the cumulative luminosity distribution implies a large, faint population, while a steep faint end implies a small, relatively bright population. Theoretical models relatively accurately predict the integrated amount of energy that is emitted by a particular population (based on time-averaged mass transfer rate and gravitational energy release), but are less accurate when it concerns the variability of sources, which relates mainly to complex accretion disk behavior.

\section{Discussion and conclusions}

We have investigated the long-term X-ray light curves collected by the \emph{RXTE} ASM for $14$ ultracompact X-ray binaries including two candidates.

All UCXBs have a significantly varying flux during their brightest phases (above $10^{37}\ \mbox{erg s}^{-1}$) as shown by the relatively flat slopes in Fig. \ref{fig:signi}. By comparing their (integrated) time-averaged bolometric luminosities with theoretical estimates (Fig. \ref{fig:avglum}), we conclude that for short-period systems (systems with an orbital period below $\sim \! 30$ min) these flat slopes probably continue at more typical luminosities (below $10^{37}\ \mbox{erg s}^{-1}$), since these relatively flat cumulative luminosity distributions are needed to match the observed and theoretical average luminosities. Longer-period systems must have a much flatter slope than short-period systems at luminosities below $10^{37}\ \mbox{erg s}^{-1}$ in order to match the theoretical models, in other words, they are more variable when faint relative to short-period systems. This follows from the steep decrease in theoretical luminosity with increasing orbital period. In general, all sources probably vary strongly at all luminosities. This confirms that long-term observations are indeed needed in order to compare a source's luminosity to a theoretical model of the mass transfer rate.

Considering the time-averaged ASM luminosity over all dwells, for the long-period systems a single power law yields an average luminosity that is too high. Instead, the slope must become flatter at lower luminosities, as already suggested by the theoretical model. The short-period systems on the other hand require a steeper low-luminosity slope.

Furthermore, several long-period systems have an average luminosity that is much higher than predicted by theory in the case of degenerate donors. A possible (partial) explanation is that the low-mass donors in these UCXBs are being heated and evaporated by irradiation from the millisecond pulsar they are orbiting. The angular momentum loss from the system resulting from a stellar wind from the donor enhances the mass transfer and hence increases the luminosity, as has been demonstrated in explaining the millisecond pulsar binary system \object{PSR J1719--1438} \citep{vanhaaften2012j1719}. A larger donor radius in itself also leads to a higher mass transfer rate.

There is no clear distinction between transient and persistent behavior in the sense that the estimated average luminosities are similar for sources over a significant range in orbital periods, which is partially a result of the similar bright-end slopes in the cumulative luminosity distributions. However, the trend that long-period systems typically have a flatter slope when they are faint compared to when they are bright can be seen as transient behavior. The variability of short-period UCXBs is more consistent with a single power-law cumulative luminosity distribution that holds up to a fraction of time of $1$, which can be understood as `persistent' behavior.

Some UCXBs seem too bright to have a degenerate donor. Especially for the long-period systems, orbital period derivatives would be very useful to decide whether the donor is helium burning or degenerate. This knowledge in turn would limit the range of theoretically predicted time-averaged luminosities, and thus constrain the faint-end luminosity distribution.

The orbital period of \object{4U 0614+091} may be much shorter than the $\sim \! 50$ min that is suggested by several observations (see Table \ref{table:params}). Both the high lower bound on the luminosity compared to the degenerate-donor model for long orbital period UCXBs, and the apparent steepening of the slope of the cumulative luminosity distribution at low luminosities, resemble the behavior of short-period systems. At least the latter argument is not conclusive, given that \object{SWIFT J1756.9--2508} combines a $54.7$ min orbital period with a steepening slope at low luminosities.

Even though the \emph{RXTE} ASM observations are the longest available, there may very well be variability on timescales (much) longer than $16$ yr, which could imply that the data for some of these systems available today are atypical. This could affect the average luminosity as derived by each of the methods, including the lower bounds. As for the long-period UCXBs, apart from being consistently brighter than theoretical estimates for systems with degenerate donors, variability on very long timescales (hundreds or thousands of years) would ensure that during any $16$ yr period, a small part of the population is exceptionally bright. This could be the population we observe.

The \emph{Monitor of All-sky X-ray Image} Gas Slit Camera (\emph{MAXI GSC}) has a higher stated sensitivity \citep{matsuoka2009} than the ASM over a shorter period of time of $2.5$ yr. Because of the much longer observation baseline, we decided to use only the ASM data. We compared the \emph{MAXI} data with the ASM data and found that the \emph{MAXI} sensitivity is better than the ASM data for 7 UCXBs with known (or tentative) orbital period, similar for 2, worse for another 2 and unavailable for 4 UCXBs. Overall the difference was not large enough to outweigh the shorter observation baseline.

We will use the luminosity distributions found in this paper in a forthcoming study of the UCXB population in the Galactic Bulge, to estimate the X-ray luminosities of systems predicted by a population model. In the context of predicting the observable population of UCXBs, the Galactic Bulge Survey \citep{jonker2011} may constrain the number of faint sources, which also gives information on the luminosity distribution.

\begin{acknowledgements}
LMvH thanks P.~G.~Jonker for explanation of ASM observations. We thank the anonymous referee for helpful comments that lead to an improvement of this paper. LMvH is supported by the Netherlands Organisation for Scientific Research (NWO). RV and GN are supported by NWO Vidi grant $016.093.305$ to GN. The \emph{RXTE} results are provided by the ASM/\emph{RXTE} teams at MIT and at the \emph{RXTE} SOF and GOF at NASA's GSFC. This research has made use of the \emph{MAXI} data provided by RIKEN, JAXA and the \emph{MAXI} team.
\end{acknowledgements}

\bibliographystyle{aa}
\bibliography{lennart_refs}

\begin{thebibliography}{93}
\expandafter\ifx\csname natexlab\endcsname\relax\def\natexlab#1{#1}\fi

\bibitem[{{Altamirano} {et~al.}(2010){Altamirano}, {Patruno}, {Heinke},
  {Markwardt}, {Strohmayer}, {Linares}, {Wijnands}, {van der Klis}, \&
  {Swank}}]{altamirano2010}
{Altamirano}, D., {Patruno}, A., {Heinke}, C.~O., {et~al.} 2010, \apjl, 712,
  L58

\bibitem[{{Bassa} {et~al.}(2006){Bassa}, {Jonker}, {in't Zand}, \&
  {Verbunt}}]{bassa2006}
{Bassa}, C.~G., {Jonker}, P.~G., {in't Zand}, J.~J.~M., \& {Verbunt}, F. 2006,
  \aap, 446, L17

\bibitem[{{Becker} {et~al.}(1977){Becker}, {Smith}, {Swank}, {Boldt}, {Holt},
  {Serlemitsos}, \& {Pravdo}}]{becker1977}
{Becker}, R.~H., {Smith}, B.~W., {Swank}, J.~H., {et~al.} 1977, \apjl, 216,
  L101

\bibitem[{{Bildsten}(1995)}]{bildsten1995}
{Bildsten}, L. 1995, \apj, 438, 852

\bibitem[{{Bradt} {et~al.}(1993){Bradt}, {Rothschild}, \& {Swank}}]{bradt1993}
{Bradt}, H.~V., {Rothschild}, R.~E., \& {Swank}, J.~H. 1993, \aaps, 97, 355

\bibitem[{{Brandt} {et~al.}(1992){Brandt}, {Castro-Tirado}, {Lund}, {Dremin},
  {Lapshov}, \& {Syunyaev}}]{brandt1992}
{Brandt}, S., {Castro-Tirado}, A.~J., {Lund}, N., {et~al.} 1992, \aap, 262, L15

\bibitem[{{Br{\'e}maud}(1997)}]{bremaud1997book}
{Br{\'e}maud}, P. 1997, {An introduction to probabilistic modeling: 3rd ed.}
  (Springer-Verlag)

\bibitem[{{Chakrabarty}(1998)}]{chakrabarty1998b}
{Chakrabarty}, D. 1998, \apj, 492, 342

\bibitem[{{Cornelisse} {et~al.}(2002){Cornelisse}, {Verbunt}, {in't Zand},
  {Kuulkers}, {Heise}, {Remillard}, {Cocchi}, {Natalucci}, {Bazzano}, \&
  {Ubertini}}]{cornelisse2002}
{Cornelisse}, R., {Verbunt}, F., {in't Zand}, J.~J.~M., {et~al.} 2002, \aap,
  392, 885

\bibitem[{{Cumming}(2003)}]{cumming2003}
{Cumming}, A. 2003, \apj, 595, 1077

\bibitem[{{Deloye} \& {Bildsten}(2003)}]{deloye2003}
{Deloye}, C.~J. \& {Bildsten}, L. 2003, \apj, 598, 1217

\bibitem[{{Deutsch} {et~al.}(2000){Deutsch}, {Margon}, \&
  {Anderson}}]{deutsch2000}
{Deutsch}, E.~W., {Margon}, B., \& {Anderson}, S.~F. 2000, \apjl, 530, L21

\bibitem[{{Dieball} {et~al.}(2005){Dieball}, {Knigge}, {Zurek}, {Shara},
  {Long}, {Charles}, {Hannikainen}, \& {van Zyl}}]{dieball2005}
{Dieball}, A., {Knigge}, C., {Zurek}, D.~R., {et~al.} 2005, \apjl, 634, L105

\bibitem[{{Dotani} {et~al.}(1990){Dotani}, {Inoue}, {Murakami}, {Nagase}, \&
  {Tanaka}}]{dotani1990}
{Dotani}, T., {Inoue}, H., {Murakami}, T., {Nagase}, F., \& {Tanaka}, Y. 1990,
  \nat, 347, 534

\bibitem[{{Dunkel} {et~al.}(2006){Dunkel}, {Chluba}, \& {Sunyaev}}]{dunkel2006}
{Dunkel}, J., {Chluba}, J., \& {Sunyaev}, R.~A. 2006, Astronomy Letters, 32,
  257

\bibitem[{{Forbes} {et~al.}(2010){Forbes}, {Evans}, {Hastings}, \&
  {Peacock}}]{forbes2010book}
{Forbes}, C., {Evans}, M., {Hastings}, N., \& {Peacock}, B. 2010, {Statistical
  Distributions: 4th ed.} (Wiley)

\bibitem[{{Ford} {et~al.}(1997){Ford}, {Kaaret}, {Tavani}, {Barret}, {Bloser},
  {Grindlay}, {Harmon}, {Paciesas}, \& {Zhang}}]{ford1997}
{Ford}, E., {Kaaret}, P., {Tavani}, M., {et~al.} 1997, \apjl, 475, L123

\bibitem[{{Forman} \& {Jones}(1976)}]{forman1976}
{Forman}, W. \& {Jones}, C. 1976, \apjl, 207, L177

\bibitem[{{Galloway} {et~al.}(2002){Galloway}, {Chakrabarty}, {Morgan}, \&
  {Remillard}}]{galloway2002}
{Galloway}, D.~K., {Chakrabarty}, D., {Morgan}, E.~H., \& {Remillard}, R.~A.
  2002, \apjl, 576, L137

\bibitem[{{Galloway} {et~al.}(2008){Galloway}, {Muno}, {Hartman}, {Psaltis}, \&
  {Chakrabarty}}]{galloway2008}
{Galloway}, D.~K., {Muno}, M.~P., {Hartman}, J.~M., {Psaltis}, D., \&
  {Chakrabarty}, D. 2008, \apjs, 179, 360

\bibitem[{{Galloway} {et~al.}(2010){Galloway}, {Yao}, {Marshall}, {Misanovic},
  \& {Weinberg}}]{galloway2010}
{Galloway}, D.~K., {Yao}, Y., {Marshall}, H., {Misanovic}, Z., \& {Weinberg},
  N. 2010, \apj, 724, 417

\bibitem[{{Grindlay} {et~al.}(1976){Grindlay}, {Gursky}, {Schnopper},
  {Parsignault}, {Heise}, {Brinkman}, \& {Schrijver}}]{grindlay1976}
{Grindlay}, J., {Gursky}, H., {Schnopper}, H., {et~al.} 1976, \apjl, 205, L127

\bibitem[{{Haggard} {et~al.}(2004){Haggard}, {Cool}, {Anderson}, {Edmonds},
  {Callanan}, {Heinke}, {Grindlay}, \& {Bailyn}}]{haggard2004}
{Haggard}, D., {Cool}, A.~M., {Anderson}, J., {et~al.} 2004, \apj, 613, 512

\bibitem[{{Hakala} {et~al.}(2011){Hakala}, {Charles}, \& {Muhli}}]{hakala2011}
{Hakala}, P.~J., {Charles}, P.~A., \& {Muhli}, P. 2011, \mnras, 416, 644

\bibitem[{{Harris}(1996)}]{harris1996}
{Harris}, W.~E. 1996, \aj, 112, 1487

\bibitem[{{Heinke} {et~al.}(2010){Heinke}, {Altamirano}, {Cohn}, {Lugger},
  {Budac}, {Servillat}, {Linares}, {Strohmayer}, {Markwardt}, {Wijnands},
  {Swank}, {Knigge}, {Bailyn}, \& {Grindlay}}]{heinke2010}
{Heinke}, C.~O., {Altamirano}, D., {Cohn}, H.~N., {et~al.} 2010, \apj, 714, 894

\bibitem[{{Heinke} {et~al.}(2001){Heinke}, {Edmonds}, \&
  {Grindlay}}]{heinke2001}
{Heinke}, C.~O., {Edmonds}, P.~D., \& {Grindlay}, J.~E. 2001, \apj, 562, 363

\bibitem[{{Hoffman} {et~al.}(1980){Hoffman}, {Cominsky}, \&
  {Lewin}}]{hoffman1980}
{Hoffman}, J.~A., {Cominsky}, L., \& {Lewin}, W.~H.~G. 1980, \apjl, 240, L27

\bibitem[{{Hoffman} {et~al.}(1976){Hoffman}, {Lewin}, {Doty}, {Hearn}, {Clark},
  {Jernigan}, \& {Li}}]{hoffman1976}
{Hoffman}, J.~A., {Lewin}, W.~H.~G., {Doty}, J., {et~al.} 1976, \apjl, 210, L13

\bibitem[{{Homer} {et~al.}(2002){Homer}, {Anderson}, {Wachter}, \&
  {Margon}}]{homer2002}
{Homer}, L., {Anderson}, S.~F., {Wachter}, S., \& {Margon}, B. 2002, \aj, 124,
  3348

\bibitem[{{Homer} {et~al.}(1996){Homer}, {Charles}, {Naylor}, {van Paradijs},
  {Auriere}, \& {Koch-Miramond}}]{homer1996}
{Homer}, L., {Charles}, P.~A., {Naylor}, T., {et~al.} 1996, \mnras, 282, L37

\bibitem[{{in 't Zand} {et~al.}(1998){in 't Zand}, {Verbunt}, {Heise},
  {Muller}, {Bazzano}, {Cocchi}, {Natalucci}, \& {Ubertini}}]{zand1998b}
{in 't Zand}, J.~J.~M., {Verbunt}, F., {Heise}, J., {et~al.} 1998, \aap, 329,
  L37

\bibitem[{{in't Zand} {et~al.}(2008){in't Zand}, {Bassa}, {Jonker}, {Keek},
  {Verbunt}, {M{\'e}ndez}, \& {Markwardt}}]{zand2008}
{in't Zand}, J.~J.~M., {Bassa}, C.~G., {Jonker}, P.~G., {et~al.} 2008, \aap,
  485, 183

\bibitem[{{in't Zand} {et~al.}(2005{\natexlab{a}}){in't Zand}, {Cornelisse}, \&
  {M{\'e}ndez}}]{zand2005a}
{in't Zand}, J.~J.~M., {Cornelisse}, R., \& {M{\'e}ndez}, M.
  2005{\natexlab{a}}, \aap, 440, 287

\bibitem[{{in't Zand} {et~al.}(2005{\natexlab{b}}){in't Zand}, {Cumming}, {van
  der Sluys}, {Verbunt}, \& {Pols}}]{zand2005}
{in't Zand}, J.~J.~M., {Cumming}, A., {van der Sluys}, M.~V., {Verbunt}, F., \&
  {Pols}, O.~R. 2005{\natexlab{b}}, \aap, 441, 675

\bibitem[{{in't Zand} {et~al.}(2007){in't Zand}, {Jonker}, \&
  {Markwardt}}]{zand2007}
{in't Zand}, J.~J.~M., {Jonker}, P.~G., \& {Markwardt}, C.~B. 2007, \aap, 465,
  953

\bibitem[{{Jonker} {et~al.}(2006){Jonker}, {Bassa}, {Nelemans}, {Juett},
  {Brown}, \& {Chakrabarty}}]{jonker2006}
{Jonker}, P.~G., {Bassa}, C.~G., {Nelemans}, G., {et~al.} 2006, \mnras, 368,
  1803

\bibitem[{{Jonker} {et~al.}(2011){Jonker}, {Bassa}, {Nelemans}, {Steeghs},
  {Torres}, {Maccarone}, {Hynes}, {Greiss}, {Clem}, {Dieball}, {Mikles},
  {Britt}, {Gossen}, {Collazzi}, {Wijnands}, {In't Zand}, {M{\'e}ndez}, {Rea},
  {Kuulkers}, {Ratti}, {van Haaften}, {Heinke}, {{\"O}zel}, {Groot}, \&
  {Verbunt}}]{jonker2011}
{Jonker}, P.~G., {Bassa}, C.~G., {Nelemans}, G., {et~al.} 2011, \apjs, 194, 18

\bibitem[{{Jonker} \& {Nelemans}(2004)}]{jonker2004}
{Jonker}, P.~G. \& {Nelemans}, G. 2004, \mnras, 354, 355

\bibitem[{{Jonker} {et~al.}(2001){Jonker}, {van der Klis}, {Homan},
  {M{\'e}ndez}, {van Paradijs}, {Belloni}, {Kouveliotou}, {Lewin}, \&
  {Ford}}]{jonker2001}
{Jonker}, P.~G., {van der Klis}, M., {Homan}, J., {et~al.} 2001, \apj, 553, 335

\bibitem[{{Juett} {et~al.}(2001){Juett}, {Psaltis}, \&
  {Chakrabarty}}]{juett2001}
{Juett}, A.~M., {Psaltis}, D., \& {Chakrabarty}, D. 2001, \apjl, 560, L59

\bibitem[{{Krimm} {et~al.}(2007){Krimm}, {Markwardt}, {Deloye}, {Romano},
  {Chakrabarty}, {Campana}, {Cummings}, {Galloway}, {Gehrels}, {Hartman},
  {Kaaret}, {Morgan}, \& {Tueller}}]{krimm2007}
{Krimm}, H.~A., {Markwardt}, C.~B., {Deloye}, C.~J., {et~al.} 2007, \apjl, 668,
  L147

\bibitem[{{Kuulkers} {et~al.}(2003){Kuulkers}, {den Hartog}, {in't Zand},
  {Verbunt}, {Harris}, \& {Cocchi}}]{kuulkers2003}
{Kuulkers}, E., {den Hartog}, P.~R., {in't Zand}, J.~J.~M., {et~al.} 2003,
  \aap, 399, 663

\bibitem[{{Kuulkers} {et~al.}(2010){Kuulkers}, {in't Zand}, {Atteia}, {Levine},
  {Brandt}, {Smith}, {Linares}, {Falanga}, {S{\'a}nchez-Fern{\'a}ndez},
  {Markwardt}, {Strohmayer}, {Cumming}, \& {Suzuki}}]{kuulkers2010}
{Kuulkers}, E., {in't Zand}, J.~J.~M., {Atteia}, J.-L., {et~al.} 2010, \aap,
  514, A65+

\bibitem[{{Landau} \& {Lifshitz}(1975)}]{landau1975}
{Landau}, L.~D. \& {Lifshitz}, E.~M. 1975, {The classical theory of fields}
  (Oxford: Pergamon Press)

\bibitem[{{Lasota}(2001)}]{lasota2001}
{Lasota}, J. 2001, \nar, 45, 449

\bibitem[{{Lasota} {et~al.}(2008){Lasota}, {Dubus}, \& {Kruk}}]{lasotaetal2008}
{Lasota}, J., {Dubus}, G., \& {Kruk}, K. 2008, \aap, 486, 523

\bibitem[{{Levine} {et~al.}(1996){Levine}, {Bradt}, {Cui}, {Jernigan},
  {Morgan}, {Remillard}, {Shirey}, \& {Smith}}]{levine1996}
{Levine}, A.~M., {Bradt}, H., {Cui}, W., {et~al.} 1996, \apjl, 469, L33

\bibitem[{{Levine} {et~al.}(2011){Levine}, {Bradt}, {Chakrabarty}, {Corbet}, \&
  {Harris}}]{levine2011}
{Levine}, A.~M., {Bradt}, H.~V., {Chakrabarty}, D., {Corbet}, R.~H.~D., \&
  {Harris}, R.~J. 2011, \apjs, 196, 6

\bibitem[{{Lewin} {et~al.}(1976){Lewin}, {Clark}, \& {Doty}}]{lewin1976}
{Lewin}, W.~H.~G., {Clark}, G., \& {Doty}, J. 1976, \iaucirc, 2922, 1

\bibitem[{{Lewin} {et~al.}(1977){Lewin}, {Hoffman}, \& {Doty}}]{lewin1977}
{Lewin}, W.~H.~G., {Hoffman}, J.~A., \& {Doty}, J. 1977, \iaucirc, 3087, 1

\bibitem[{{Liu} {et~al.}(2007){Liu}, {van Paradijs}, \& {van den
  Heuvel}}]{liu2007}
{Liu}, Q.~Z., {van Paradijs}, J., \& {van den Heuvel}, E.~P.~J. 2007, \aap,
  469, 807

\bibitem[{{Madej} \& {Jonker}(2011)}]{madej2011}
{Madej}, O.~K. \& {Jonker}, P.~G. 2011, \mnras, 412, L11

\bibitem[{{Madej} {et~al.}(2010){Madej}, {Jonker}, {Fabian}, {Pinto},
  {Verbunt}, \& {de Plaa}}]{madej2010}
{Madej}, O.~K., {Jonker}, P.~G., {Fabian}, A.~C., {et~al.} 2010, \mnras, 407,
  L11

\bibitem[{{Markwardt} {et~al.}(2003{\natexlab{a}}){Markwardt}, {Juda}, \&
  {Swank}}]{markwardt2003b}
{Markwardt}, C.~B., {Juda}, M., \& {Swank}, J.~H. 2003{\natexlab{a}}, ATel,
  127, 1

\bibitem[{{Markwardt} {et~al.}(2003{\natexlab{b}}){Markwardt}, {Smith}, \&
  {Swank}}]{markwardt2003c}
{Markwardt}, C.~B., {Smith}, E., \& {Swank}, J.~H. 2003{\natexlab{b}},
  \iaucirc, 8080, 2

\bibitem[{{Markwardt} {et~al.}(2002){Markwardt}, {Swank}, {Strohmayer}, {in 't
  Zand}, \& {Marshall}}]{markwardt2002}
{Markwardt}, C.~B., {Swank}, J.~H., {Strohmayer}, T.~E., {in 't Zand},
  J.~J.~M., \& {Marshall}, F.~E. 2002, \apjl, 575, L21

\bibitem[{{Matsuoka} {et~al.}(2009){Matsuoka}, {Kawasaki}, {Ueno}, {Tomida},
  {Kohama}, {Suzuki}, {Adachi}, {Ishikawa}, {Mihara}, {Sugizaki}, {Isobe},
  {Nakagawa}, {Tsunemi}, {Miyata}, {Kawai}, {Kataoka}, {Morii}, {Yoshida},
  {Negoro}, {Nakajima}, {Ueda}, {Chujo}, {Yamaoka}, {Yamazaki}, {Nakahira},
  {You}, {Ishiwata}, {Miyoshi}, {Eguchi}, {Hiroi}, {Katayama}, \&
  {Ebisawa}}]{matsuoka2009}
{Matsuoka}, M., {Kawasaki}, K., {Ueno}, S., {et~al.} 2009, \pasj, 61, 999

\bibitem[{{Middleditch} {et~al.}(1981){Middleditch}, {Mason}, {Nelson}, \&
  {White}}]{middleditch1981}
{Middleditch}, J., {Mason}, K.~O., {Nelson}, J.~E., \& {White}, N.~E. 1981,
  \apj, 244, 1001

\bibitem[{{Nelemans} \& {Jonker}(2010)}]{nelemans2010b}
{Nelemans}, G. \& {Jonker}, P.~G. 2010, \nar, 54, 87

\bibitem[{{Nelemans} {et~al.}(2004){Nelemans}, {Jonker}, {Marsh}, \& {van der
  Klis}}]{nelemans2004}
{Nelemans}, G., {Jonker}, P.~G., {Marsh}, T.~R., \& {van der Klis}, M. 2004,
  \mnras, 348, L7

\bibitem[{{Nelemans} {et~al.}(2006){Nelemans}, {Jonker}, \&
  {Steeghs}}]{nelemans2006}
{Nelemans}, G., {Jonker}, P.~G., \& {Steeghs}, D. 2006, \mnras, 370, 255

\bibitem[{{Nelson} {et~al.}(1986){Nelson}, {Rappaport}, \& {Joss}}]{nelson1986}
{Nelson}, L.~A., {Rappaport}, S.~A., \& {Joss}, P.~C. 1986, \apj, 304, 231

\bibitem[{{Ortolani} {et~al.}(1994){Ortolani}, {Barbuy}, \&
  {Bica}}]{ortolani1994}
{Ortolani}, S., {Barbuy}, B., \& {Bica}, E. 1994, \aaps, 108, 653

\bibitem[{{Osaki}(1974)}]{osaki1974}
{Osaki}, Y. 1974, \pasj, 26, 429

\bibitem[{{Paczy{\'n}ski} \& {Sienkiewicz}(1981)}]{paczynski1981}
{Paczy{\'n}ski}, B. \& {Sienkiewicz}, R. 1981, \apjl, 248, L27

\bibitem[{{Podsiadlowski} {et~al.}(2002){Podsiadlowski}, {Rappaport}, \&
  {Pfahl}}]{podsiadlowski2002}
{Podsiadlowski}, P., {Rappaport}, S., \& {Pfahl}, E.~D. 2002, \apj, 565, 1107

\bibitem[{{Press} {et~al.}(1992){Press}, {Teukolsky}, {Vetterling}, \&
  {Flannery}}]{press1992book}
{Press}, W.~H., {Teukolsky}, S.~A., {Vetterling}, W.~T., \& {Flannery}, B.~P.
  1992, {Numerical Recipes in Fortran 77: 2nd ed.} (Cambridge, UK: Cambridge
  University Press)

\bibitem[{{Rappaport} {et~al.}(1983){Rappaport}, {Verbunt}, \&
  {Joss}}]{rappaport1983}
{Rappaport}, S., {Verbunt}, F., \& {Joss}, P.~C. 1983, \apj, 275, 713

\bibitem[{{Savonije} {et~al.}(1986){Savonije}, {de Kool}, \& {van den
  Heuvel}}]{savonije1986}
{Savonije}, G.~J., {de Kool}, M., \& {van den Heuvel}, E.~P.~J. 1986, \aap,
  155, 51

\bibitem[{{Schulz} {et~al.}(2001){Schulz}, {Chakrabarty}, {Marshall},
  {Canizares}, {Lee}, \& {Houck}}]{schulz2001}
{Schulz}, N.~S., {Chakrabarty}, D., {Marshall}, H.~L., {et~al.} 2001, \apj,
  563, 941

\bibitem[{{Shahbaz} {et~al.}(2008){Shahbaz}, {Watson}, {Zurita}, {Villaver}, \&
  {Hernandez-Peralta}}]{shahbaz2008}
{Shahbaz}, T., {Watson}, C.~A., {Zurita}, C., {Villaver}, E., \&
  {Hernandez-Peralta}, H. 2008, \pasp, 120, 848

\bibitem[{{Smak}(1984)}]{smak1984}
{Smak}, J. 1984, \pasp, 96, 5

\bibitem[{{Stella} {et~al.}(1987){Stella}, {Priedhorsky}, \&
  {White}}]{stella1987}
{Stella}, L., {Priedhorsky}, W., \& {White}, N.~E. 1987, \apjl, 312, L17

\bibitem[{{Strohmayer} {et~al.}(2008){Strohmayer}, {Markwardt}, \&
  {Kuulkers}}]{strohmayer2008}
{Strohmayer}, T.~E., {Markwardt}, C.~B., \& {Kuulkers}, E. 2008, \apjl, 672,
  L37

\bibitem[{{Strohmayer} {et~al.}(1996){Strohmayer}, {Zhang}, {Swank}, {Smale},
  {Titarchuk}, {Day}, \& {Lee}}]{strohmayer1996}
{Strohmayer}, T.~E., {Zhang}, W., {Swank}, J.~H., {et~al.} 1996, \apjl, 469, L9

\bibitem[{{Swank} {et~al.}(1976){Swank}, {Becker}, {Pravdo}, {Saba}, \&
  {Serlemitsos}}]{swank1976}
{Swank}, J.~H., {Becker}, R.~H., {Pravdo}, S.~H., {Saba}, J.~R., \&
  {Serlemitsos}, P.~J. 1976, \iaucirc, 3010, 1

\bibitem[{{Swank} {et~al.}(1978){Swank}, {Boldt}, {Holt}, {Serlemitsos}, \&
  {Becker}}]{swank1978}
{Swank}, J.~H., {Boldt}, E.~A., {Holt}, S.~S., {Serlemitsos}, P.~J., \&
  {Becker}, R.~H. 1978, \mnras, 182, 349

\bibitem[{{Tan} {et~al.}(1991){Tan}, {Morgan}, {Lewin}, {Penninx}, {van der
  Klis}, {van Paradijs}, {Makishima}, {Inoue}, {Dotani}, \&
  {Mitsuda}}]{tan1991}
{Tan}, J., {Morgan}, E., {Lewin}, W.~H.~G., {et~al.} 1991, \apj, 374, 291

\bibitem[{{van der Sluys} {et~al.}(2005){van der Sluys}, {Verbunt}, \&
  {Pols}}]{sluys2005a}
{van der Sluys}, M.~V., {Verbunt}, F., \& {Pols}, O.~R. 2005, \aap, 431, 647

\bibitem[{{van Haaften} {et~al.}(2012{\natexlab{a}}){van Haaften}, {Nelemans},
  {Voss}, \& {Jonker}}]{vanhaaften2012j1719}
{van Haaften}, L.~M., {Nelemans}, G., {Voss}, R., \& {Jonker}, P.~G.
  2012{\natexlab{a}}, \aap, 541, A22

\bibitem[{{van Haaften} {et~al.}(2012{\natexlab{b}}){van Haaften}, {Nelemans},
  {Voss}, {Wood}, \& {Kuijpers}}]{vanhaaften2012}
{van Haaften}, L.~M., {Nelemans}, G., {Voss}, R., {Wood}, M.~A., \& {Kuijpers},
  J. 2012{\natexlab{b}}, \aap, 537, A104

\bibitem[{{van Paradijs} {et~al.}(1990){van Paradijs}, {Dotani}, {Tanaka}, \&
  {Tsuru}}]{vanparadijs1990}
{van Paradijs}, J., {Dotani}, T., {Tanaka}, Y., \& {Tsuru}, T. 1990, \pasj, 42,
  633

\bibitem[{{Walter} {et~al.}(1982){Walter}, {Mason}, {Clarke}, {Halpern},
  {Grindlay}, {Bowyer}, \& {Henry}}]{walter1982}
{Walter}, F.~M., {Mason}, K.~O., {Clarke}, J.~T., {et~al.} 1982, \apjl, 253,
  L67

\bibitem[{{Wang} \& {Chakrabarty}(2004)}]{wang2004}
{Wang}, Z. \& {Chakrabarty}, D. 2004, \apjl, 616, L139

\bibitem[{{Werner} {et~al.}(2006){Werner}, {Nagel}, {Rauch}, {Hammer}, \&
  {Dreizler}}]{werner2006}
{Werner}, K., {Nagel}, T., {Rauch}, T., {Hammer}, N.~J., \& {Dreizler}, S.
  2006, \aap, 450, 725

\bibitem[{{White} \& {Swank}(1982)}]{white1982}
{White}, N.~E. \& {Swank}, J.~H. 1982, \apjl, 253, L61

\bibitem[{{Wijnands} {et~al.}(2005){Wijnands}, {Homan}, {Heinke}, {Miller}, \&
  {Lewin}}]{wijnands2005}
{Wijnands}, R., {Homan}, J., {Heinke}, C.~O., {Miller}, J.~M., \& {Lewin},
  W.~H.~G. 2005, \apj, 619, 492

\bibitem[{{Wilson-Hodge} {et~al.}(2012){Wilson-Hodge}, {Case}, {Cherry},
  {Rodi}, {Camero-Arranz}, {Jenke}, {Chaplin}, {Beklen}, {Finger}, {Bhat},
  {Briggs}, {Connaughton}, {Greiner}, {Kippen}, {Meegan}, {Paciesas}, {Preece},
  \& {von Kienlin}}]{wilsonhodge2012arXiv}
{Wilson-Hodge}, C.~A., {Case}, G.~L., {Cherry}, M.~L., {et~al.} 2012, ArXiv
  e-prints, 1201.3585

\bibitem[{{Yungelson}(2008)}]{yungelson2008}
{Yungelson}, L.~R. 2008, Astronomy Letters, 34, 620

\bibitem[{{Zhang} {et~al.}(2012){Zhang}, {Hynes}, \& {Robinson}}]{zhang2012}
{Zhang}, Y., {Hynes}, R.~I., \& {Robinson}, E.~L. 2012, \mnras, 419, 2943

\bibitem[{{Zhong} \& {Wang}(2011)}]{zhong2011}
{Zhong}, J. \& {Wang}, Z. 2011, \apj, 729, 8

\bibitem[{{Zurek} {et~al.}(2009){Zurek}, {Knigge}, {Maccarone}, {Dieball}, \&
  {Long}}]{zurek2009}
{Zurek}, D.~R., {Knigge}, C., {Maccarone}, T.~J., {Dieball}, A., \& {Long},
  K.~S. 2009, \apj, 699, 1113

\end{thebibliography}

\end{document}